\documentclass[pdflatex,sn-mathphys-num]{sn-jnl}

\usepackage{graphicx}%
\usepackage{multirow}%
\usepackage{amsmath,amssymb,amsfonts}%
\usepackage{amsthm}%
\usepackage{mathrsfs}%
\usepackage[title]{appendix}%
\usepackage{xcolor}%
\usepackage{textcomp}%
\usepackage{manyfoot}%
\usepackage{booktabs}%
\usepackage{algorithm}%
\usepackage{algorithmicx}%
\usepackage{algpseudocode}%
\usepackage{listings}%

\usepackage{subfig}
\usepackage{lineno}
\usepackage{threeparttable} 

\usepackage{xcolor}
\usepackage{longtable}
\def\BibTeX{{\rm B\kern-.05em{\sc i\kern-.025em b}\kern-.08em
    T\kern-.1667em\lower.7ex\hbox{E}\kern-.125emX}}

\theoremstyle{thmstyleone}%
\theoremstyle{thmstyletwo}%

\theoremstyle{thmstylethree}%

\begin{document}

\title[tcrLM]{tcrLM: a lightweight protein language model for predicting T cell receptor and epitope binding specificity}


\author[1]{\fnm{Chenpeng} \sur{Yu}}

\author[1]{\fnm{Xing} \sur{Fang}}

\author[1]{\fnm{Shiye} \sur{Tian}}

\author*[1]{\fnm{Hui} \sur{Liu}}\email{hliu@njtech.edu.cn}

\affil*[1]{\orgdiv{College of Computer and Information Engineering}, \orgname{Nanjing Tech University}, \orgaddress{\city{Nanjing}, \postcode{211800},  \country{China}}}




\abstract{
The anti-cancer immune response relies on the bindings between T-cell receptors (TCRs) and antigens, which elicits adaptive immunity to eliminate tumor cells. This ability of the immune system to respond to novel various neoantigens arises from the immense diversity of TCR repository. However, TCR diversity poses a significant challenge on accurately predicting antigen-TCR bindings. In this study, we introduce a lightweight masked language model, termed tcrLM, to address this challenge. Our approach involves randomly masking segments of TCR sequences and training tcrLM to infer the masked segments, thereby enabling the extraction of expressive features from TCR sequences. To further enhance robustness, we incorporate virtual adversarial training into tcrLM. We construct the largest TCR CDR3 sequence set with more than 100 million distinct sequences, and pretrain tcrLM on these sequences. The pre-trained encoder is subsequently applied to predict TCR-antigen binding specificity. We evaluate model performance on three test datasets: independent, external, and COVID-19 test set. The results demonstrate that tcrLM not only surpasses existing TCR-antigen binding prediction methods, but also outperforms other mainstream protein language models. More interestingly, tcrLM effectively captures the biochemical properties and positional preference of amino acids within TCR sequences. Additionally, the predicted TCR-neoantigen binding scores indicates the immunotherapy responses and clinical outcomes in a melanoma cohort. These findings demonstrate the potential of tcrLM in predicting TCR-antigen binding specificity, with significant implications for advancing immunotherapy and personalized medicine.  }




\maketitle

\section{Introduction}\label{sec1}
The recognition of antigens is a crucial step in initiating T cell-mediated immune responses against cancer~\cite{glanville2017identifying,zhang2019combination}. This process is orchestrated by the interactions between T cells and antigen-presenting cells (APCs)~\cite{yang2023antigen}, followed by the clonal expansion and differentiation of naive T cells into effector T cells, and their migration to the tumor site to eliminate cancer cells~\cite{chen2013oncology}. The T-cell receptor (TCR) specifically recognizes antigens presented by major histocompatibility complex (MHC) class I molecules on the surface of APCs. This recognition occurs through binding to the antigen-MHC complex, triggering a series of intracellular signaling events that result in T cell activation. Upon activation, T cells undergo clonal expansion and differentiate into various effector subsets, most notably cytotoxic T lymphocytes (CTLs)~\cite{barry2002cytotoxic,raskov2021cytotoxic,weigelin2021cytotoxic}, which are central to the anti-tumor immune response by directly targeting and eliminating tumor cells. Consequently, the peptide-TCR binding (pTCR) represents a pivotal step in the immune response process~\cite{yewdell1999immunodominance,dunn2004three}.

The TCR comprises $\alpha$ and $\beta$ chains that are generated through genetic recombination, leading to an expansive TCR repertoire. Previous studies have documented that human has the potential to generate approximately 10$^{15}$ to 10$^{20}$ distinct TCR sequences. This diversity predominantly manifests in the complementarity determining region 3 (CDR3)~\cite{zhang2016direct}, which engages directly with the peptide-MHC complex, thereby dictating the binding specificity of the TCR \cite{davis1988t,krogsgaard2005t}. Quite a few computational methods have been developed to predict the binding specificity of T-cell receptors (TCRs) to peptide-MHC complexes (pMHCs). These methods roughly fall into three main categories: 1) Clustering-based methods measure the similarities between TCRs and try to understand underlying patterns in binding to antigens. Representative methods include TCRdist~\cite{dash2017quantifiable}, DeepTCR~\cite{sidhom2021deeptcr}, GIANA~\cite{zhang2021giana}, iSMART~\cite{zhang2020investigation},  GLIPH~\cite{glanville2017identifying}, and ELATE~\cite{dvorkin2021autoencoder} models. 2) Peptide-specific models focus on predicting the binding of specific peptide to TCRs, including TCRGP~\cite{jokinen2021predicting}, TCRex~\cite{gielis2019detection}, and NetTCR-2~\cite{montemurro2021nettcr}. 3) Generic prediction models are not limited to specific peptides but need training on known pTCR bindings, including PanPep~\cite{gao2023pan}, pMTnet~\cite{lu2021deep}, DLpTCR~\cite{xu2021dlptcr}, ERGO2~\cite{springer2020prediction}, and TITAN~\cite{weber2021titan}. Although these methods have demonstrated promising accuracy in specific scenarios, they remain limited in generalizing to unseen peptides, which is crucial for identifying the binding specificity of tumor neoantigens or exogenous antigens. Therefore, accurately identifying the pTCR bindings remains a challenging task. 

Essentially, both the TCR CDR3 and antigens can be expressed as a sequence of amino acids, exhibiting striking similarity to human natural language. Large language models have gained increasing attention in the field of protein modeling in recent years~\cite{devlin2018bert,radford2018improving,brown2020language}, and yield to new insights into the understanding for structures and functions of proteins. Most protein language models are built upon the Transformer encoder\cite{vaswani2017attention}, which can encode protein sequences or structures into predefined-length latent representations. The representations have been proven to boost the performance of downstream tasks related to proteins, such as the prediction of structures, functions, and protein-drug interactions. The masked language models (MLM), such as CVC~\cite{goldner2024self}, TCR-BERT~\cite{wu2024tcr}, ESM1b~\cite{rives2021biological}, ESM-1v~\cite{meier2021language}, ProtFlash~\cite{wang2023deciphering} and ProtTrans~\cite{elnaggar2021prottrans}, aim to reconstruct masked tokens based on surrounding sequences. Their success has inspired to make advantage of MLM to accommodate the diversity of TCR sequences. For example, TULIP~\cite{meynard2024tulip} exploits the Transformer-based unsupervised language model to predict the interactions between peptides and TCRs. TAPIR~\cite{fast2023tapir} is also inspired by language models and involves a two-tower architecture composed of independent CNN-based encoders for TCR and target sequences. ProtLM.TCR~\cite{essaghir2022t} is a RoBERTa-style Transformer-based language model designed to predict binding between TCR and HLA class I epitope sequences. CVC~\cite{ goldner2024self} uses a lightly modified BERT architecture with tweaked pre-training objectives trained on TCR CDR3 sequences. TCR-BERT~\cite{wu2024tcr} adopts self-supervised learning to learn latent representations of TCR CDR3 sequences, thereby enhancing the performance of downstream tasks, such as TCR-antigen binding prediction, sequence clustering, and engineered TCR design.

In this study, we introduce tcrLM, a lightweight masked language model designed to predict TCR-antigen binding specificity. By randomly masking segments of TCR sequences and training it to infer the masked regions, we extract informative and expressive features from TCR sequences. To enhance model robustness, we incorporate virtual adversarial training into tcrLM architecture. For pretraining, we curated the largest set of TCR CDR3 sequences to date (113,888,692 distinct sequences). After pretraining, we fine-tune it for TCR-antigen binding prediction. The model was evaluated on three test datasets—independent, external, and COVID-19—and the results verified that tcrLM outperformed existing TCR prediction methods as well as mainstream protein language models. Furthermore, we validated the correlation between the tcrLM’s predicted TCR-neoantigen binding scores and immunotherapy outcomes in a melanoma cohort. Interestingly, tcrLM was able to effectively captures the biochemical properties and positional preference of amino acids within the TCR CDR3 sequences. These findings underscore the potential of tcrLM in advancing immunotherapy and personalized medicine. In summary, we believe that this study makes at least four significant contributions: 1) We develop a lightweight masked language model to learn the representations of TCR CDR3 sequence. Also, virtual adversarial training is introduced to reduce the model's sensitivity to slight input variations, thereby enhance its generalizability. 2) For pretraining, we curated more than 100 million distinct TCR CDR3 sequences from more than ten databases and publications. To our best knowledge, this is the largest set of TCR CDR3 sequence to date. Also, we established three large-scale pTCR binding test sets for performance evaluation. 3) Our extensive experiments have demonstrated that tcrLM outperforms current state-of-the-art pTCR binding prediction methods, and also exhibits superior performance compared to other protein language models, even those with significantly larger parameter sizes. 4) We demonstrate that the predicted pTCR binding scores closely correlate to the immunotherapy outcomes in a melanoma cohort. Moreover, the attention scores enable us to capture residual biochemical properties and positional preference.

\section{Results}
\subsection{Overview of TCR language model}
\begin{figure*}[!htbp]
    \centering
    \begin{minipage}{1\textwidth}
        \centering
        \subfloat{\includegraphics[width=1\textwidth]{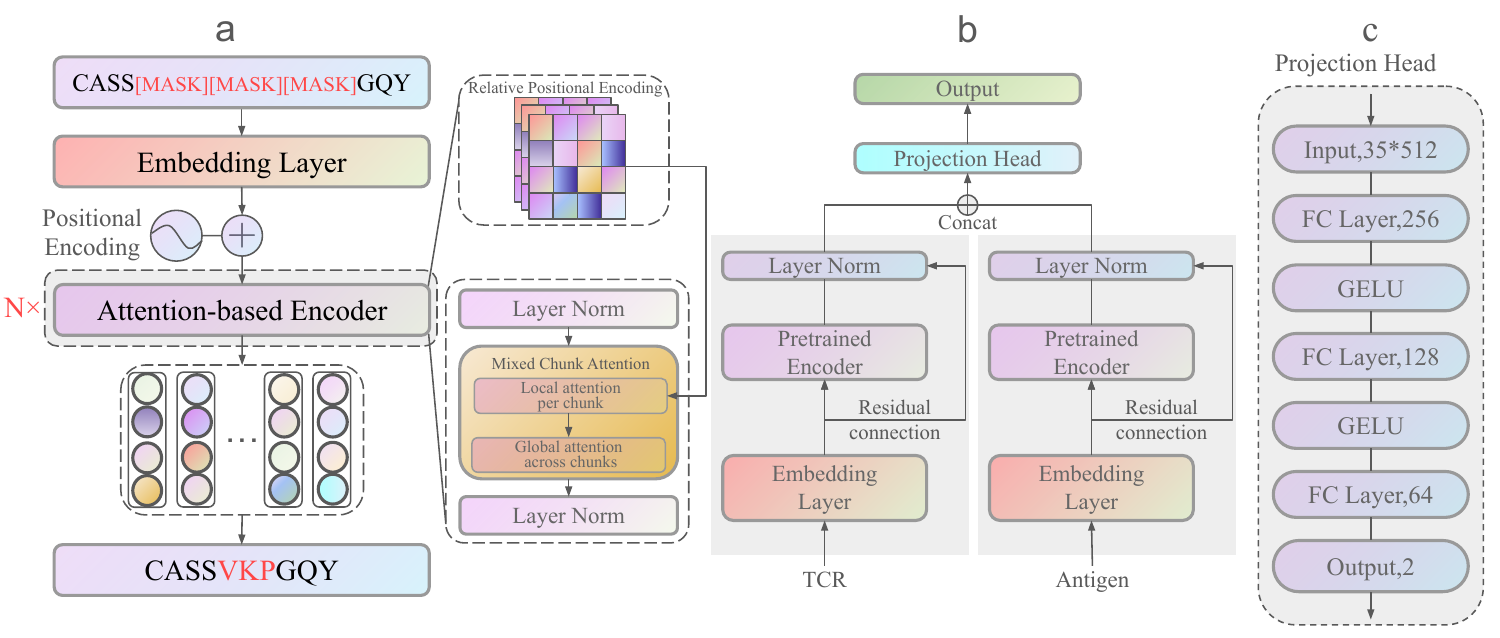}}
    \end{minipage}
    \hfill
    \caption{Illustrative diagram of tcrLM model. (a) Masked language model with the encoder composed of stacked attention-based modules. (b) Prediction network includes the pre-trained encoder to generate embeddings and projection head for pTCR binding prediction. (c) Detailed architecture of the projection head.}
    \label{fig:model}
\end{figure*}

We utilize the BERT-based language model~\cite{devlin2018bert} and its training methodology to effectively process and analyze TCR sequences (Fig.~\ref{fig:model}). Our model employs a Masked Language Model to learn bidirectional contextual information by reconstructing masked tokens from their surrounding context. The encoder, composed of stacked attention mechanism modules, is trained to capture meaningful representations of TCR sequences. Unlike other protein language models trained on generic protein sequences, such as UniRef50 or UniRef90 set, we compile a dataset comprising over 100 million TCR CDR3 distinct sequences from diverse sources to pre-train our model. To our best knowledge, this is the largest collection of human TCR CDR3 sequences to date. During the fine-tuning stage, the pre-trained encoder is used to derive the embeddings of TCR and antigen sequences, which are subsequently concatenated and passed through a projection head to predict pTCR binding specificity.

\subsection{Performance evaluation on independent test set}
We first assessed the performance of tcrLM on the 10\% hold-out independent test set. To benchmark the performance of our method, we conducted a comparative analysis with five current state-of-the-art methods: PanPep~\cite{gao2023pan}, ERGO2~\cite{springer2020prediction}, pMTnet ~\cite{lu2021deep}, and DLpTCR~\cite{xu2021dlptcr}, as well as a TCR language model TAPIR ~\cite{fast2023tapir}. For the former three methods, we executed their executable codes using their recommended parameters on the same workstation to tcrLM. For DLpTCR and TAPIR, we utilized their web servers to make predictions on the test set.

\begin{figure*}[!htbp]
    \centering
    \begin{minipage}[t]{0.49\linewidth}
        \centering
        \subfloat[ROC curves]{\includegraphics[width=\linewidth]{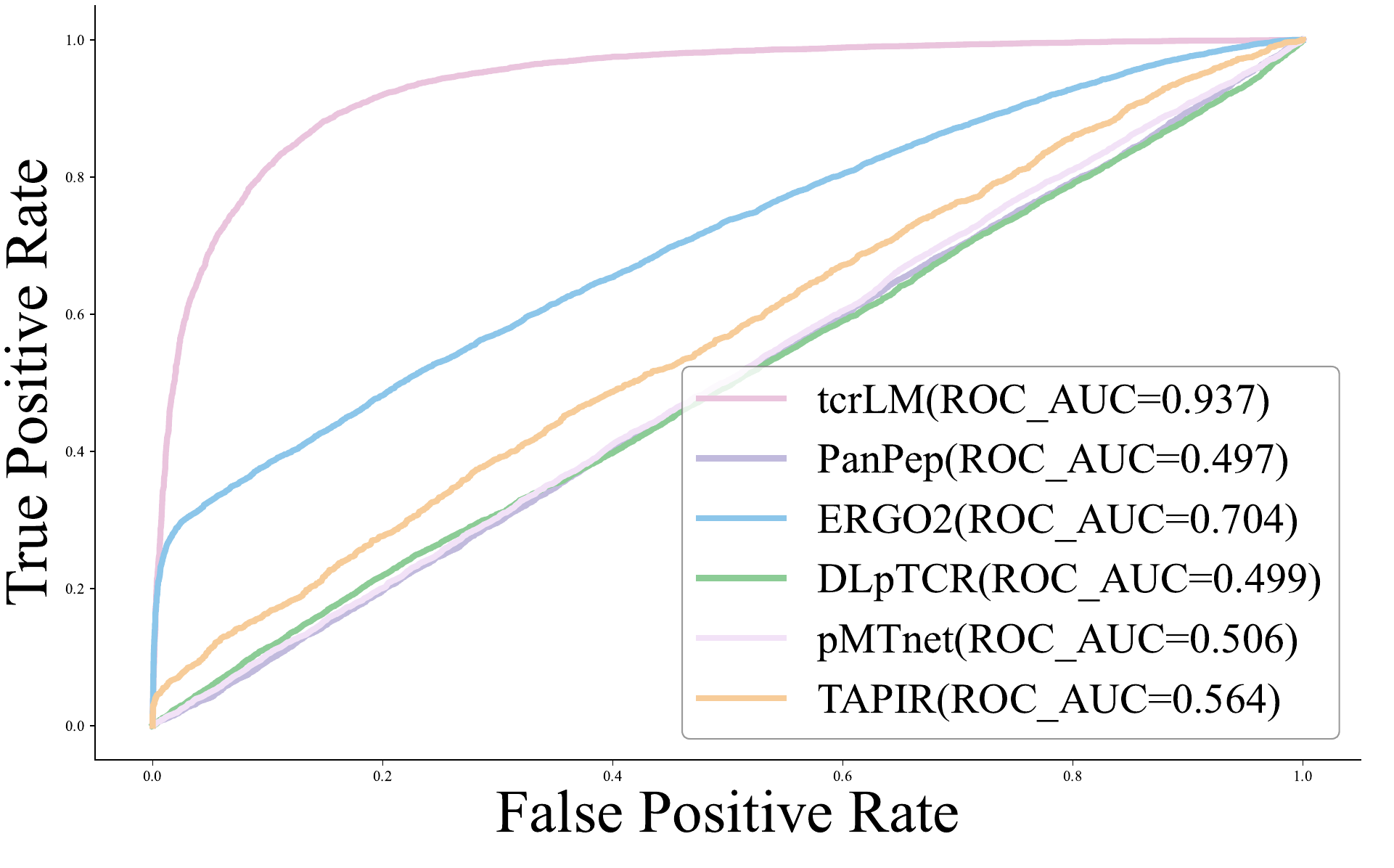}}
    \end{minipage}
        \begin{minipage}[t]{0.49\linewidth}
        \centering
        \subfloat[PR curves]{\includegraphics[width=\linewidth]{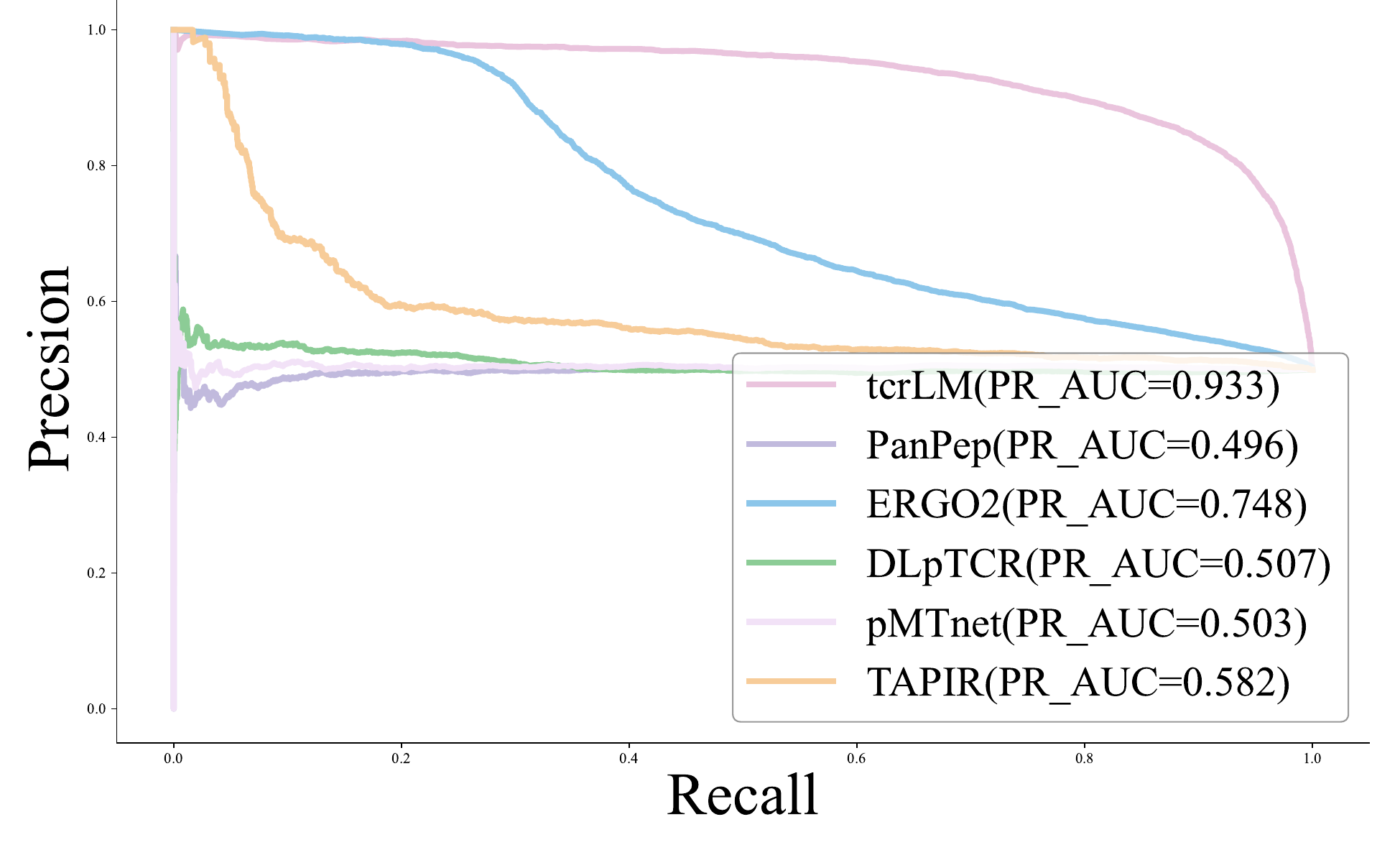}}
    \end{minipage} \\
    \vspace{0.3cm}
    \begin{minipage}[t]{1\linewidth}
        \centering
        \subfloat[Performance comparison]{\includegraphics[width=0.465\linewidth]{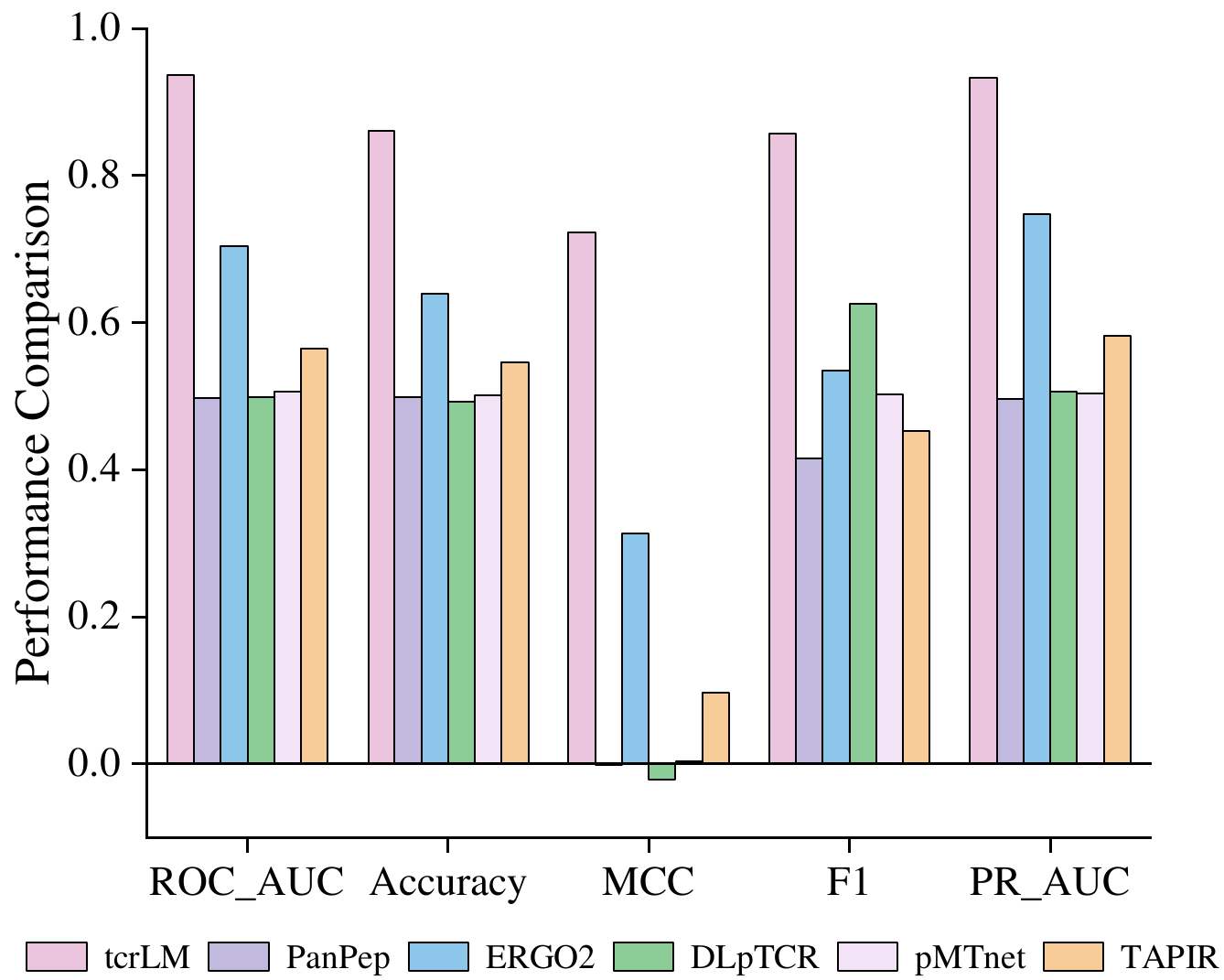}} 
        \subfloat[PPV metric]{\includegraphics[width=0.535\linewidth]{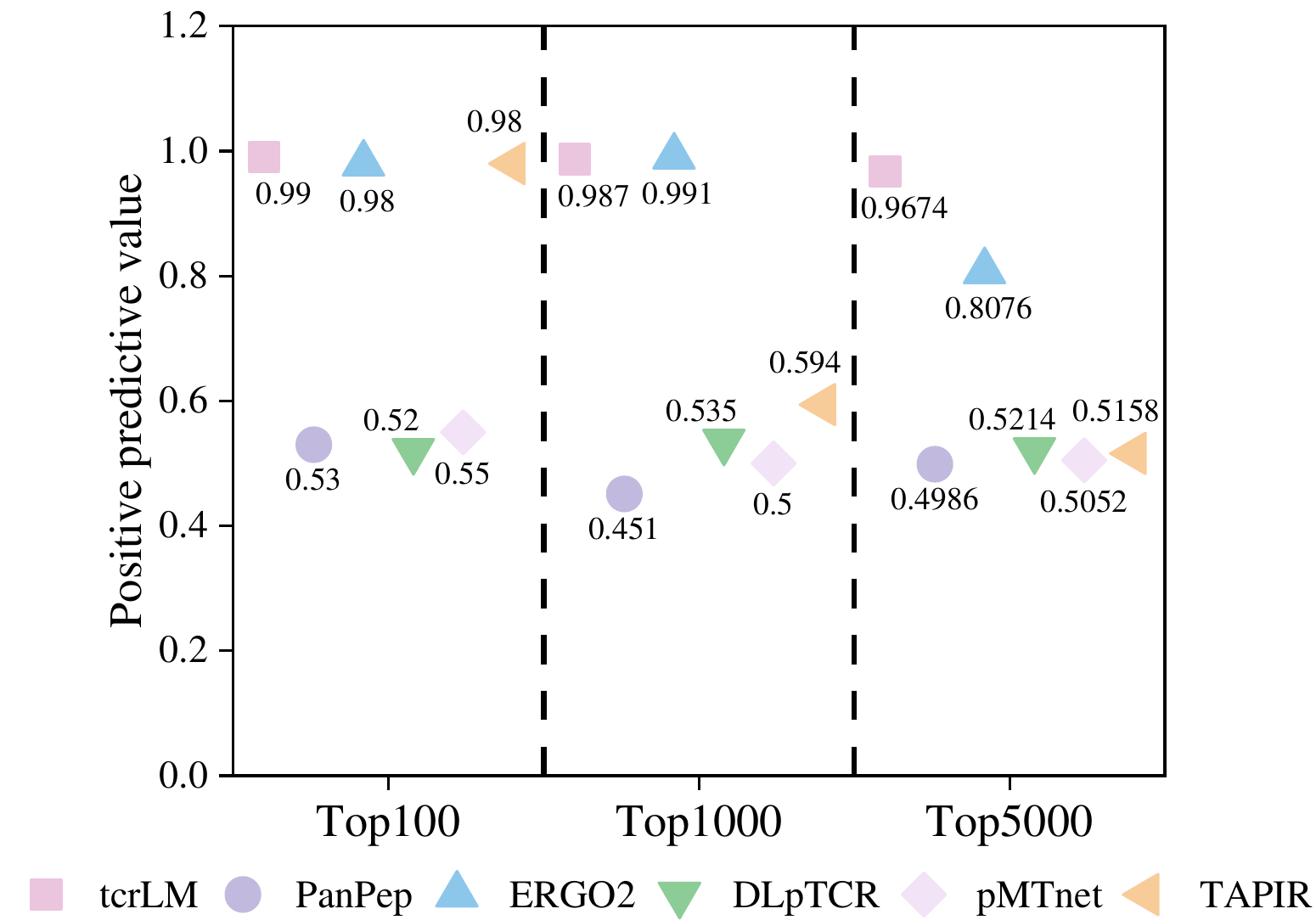}}
    \end{minipage} 

        \vspace{0.3cm}
    \begin{minipage}[t]{\linewidth}
        \centering
        \subfloat[Performance evaluation of ablated models]{\includegraphics[width=\linewidth]{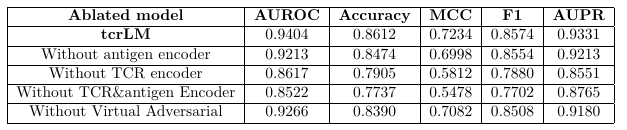}}
    \end{minipage}
    \caption{Performance evaluation of tcrLM and five comparative methods on independent test set. (a-b) The ROC and Precision-Recall curves achieved by tcrLM and competing methods on the independent test set, respectively. (c) Comprehensive performance comparison between tcrLM and five competing methods. (d) Positive predictive value (PPV) for the top 100, top 1000, and top 5000 predictions. (e) Model ablation experiments for verifying the effectiveness of pretrained encoder and virtual adversarial training. }
    \label{fig:independent}
\end{figure*}

We found that tcrLM significantly outperforms all other methods ( Fig.~\ref{fig:independent}a-c). Specifically, tcrLM achieved an AUROC of 0.937 and an AUPR of 0.933, highlighting its exceptional predictive capability in determining pTCR binding specificity. Among the compared methods, only ERGO2 demonstrated moderate performance, with an AUROC of 0.704 and an AUPR of 0.747, while the remaining methods performed close to random guessing. The existing methods may showed promising performance on small test set, but their performance declined significantly when tested on the relatively large test set. This suggests that they have weak generalizability and struggle to adapt to large-scale test data scenarios. To further examine our model's ability to prioritize pTCR bindings, we calculated the positive predictive value (PPV) for the top-ranked predicted pTCR samples. Specifically, we computed the PPV for the top 100, top 1000, and top 5000 predictions. The results showed that tcrLM achieved PPV values of 99\%, 98.7\%, and 96.74\% for the top 100, top 1000, and top 5000 predictions (Fig.~\ref{fig:independent}d), respectively. In comparison, other methods demonstrated inferior prioritization capacity compared to tcrLM.

To verify the effectiveness of the pre-trained encoder, we conducted ablation experiments to evaluate its impact on the performance of pTCR prediction. Instead of utilizing the pretrained encoder to extract features from input sequences, we directly fed the sequence embeddings into the projection head for pTCR binding prediction. This approach allowed us to independently assess the impact of removing the TCR encoder, the antigen encoder, or both simultaneously. The performance of the three ablated models revealed that removal of the pre-trained encoder consistently led to performance decline (Fig.~\ref{fig:independent}e). In particular, the removal of TCR encoder resulted in a notable drop of AUC value from 0.94 to 0.86, highlighting the importance of the pre-trained encoder for feature extraction of TCR sequence. We also tested the effect of virtual adversarial training, and found that removal of virtual adversarial training led to about a 2\% performance reduction.

\subsection{Evaluation on external and COVID-19 test dataset}
\begin{figure*}[!htbp]
    \centering
    \begin{minipage}[t]{0.49\linewidth}
        \centering
         \subfloat[Performance on external set]{\includegraphics[width=0.9\linewidth]{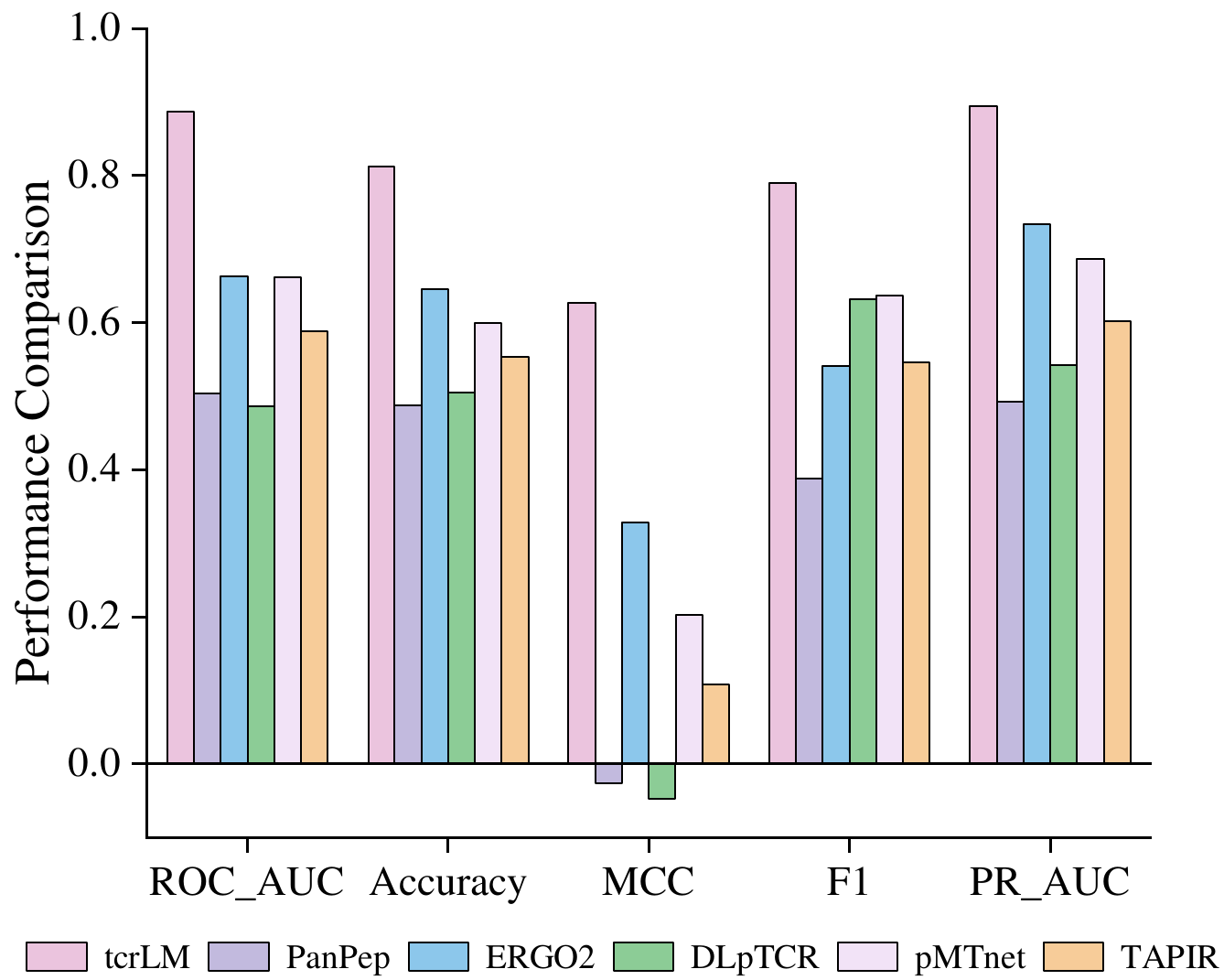}}
    \end{minipage}
    \begin{minipage}[t]{0.49\linewidth}
        \centering
         \subfloat[Performance on COVID-19 set]{\includegraphics[width=0.9\linewidth]{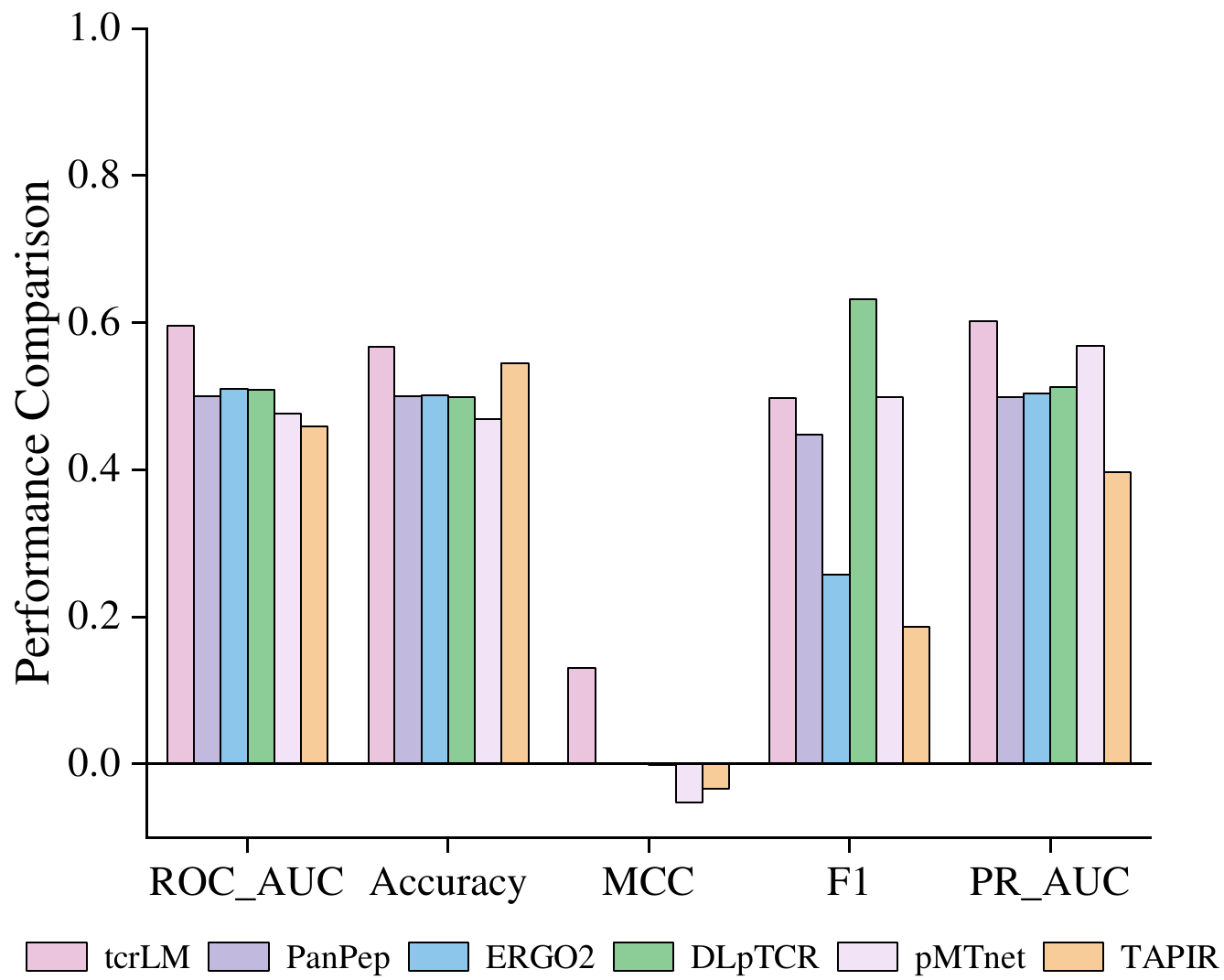}}
    \end{minipage} \\
    \vspace{0.5cm}
    \begin{minipage}[t]{1\linewidth}
        \centering
         \subfloat[PPV on external set]{\includegraphics[width=0.5\linewidth]{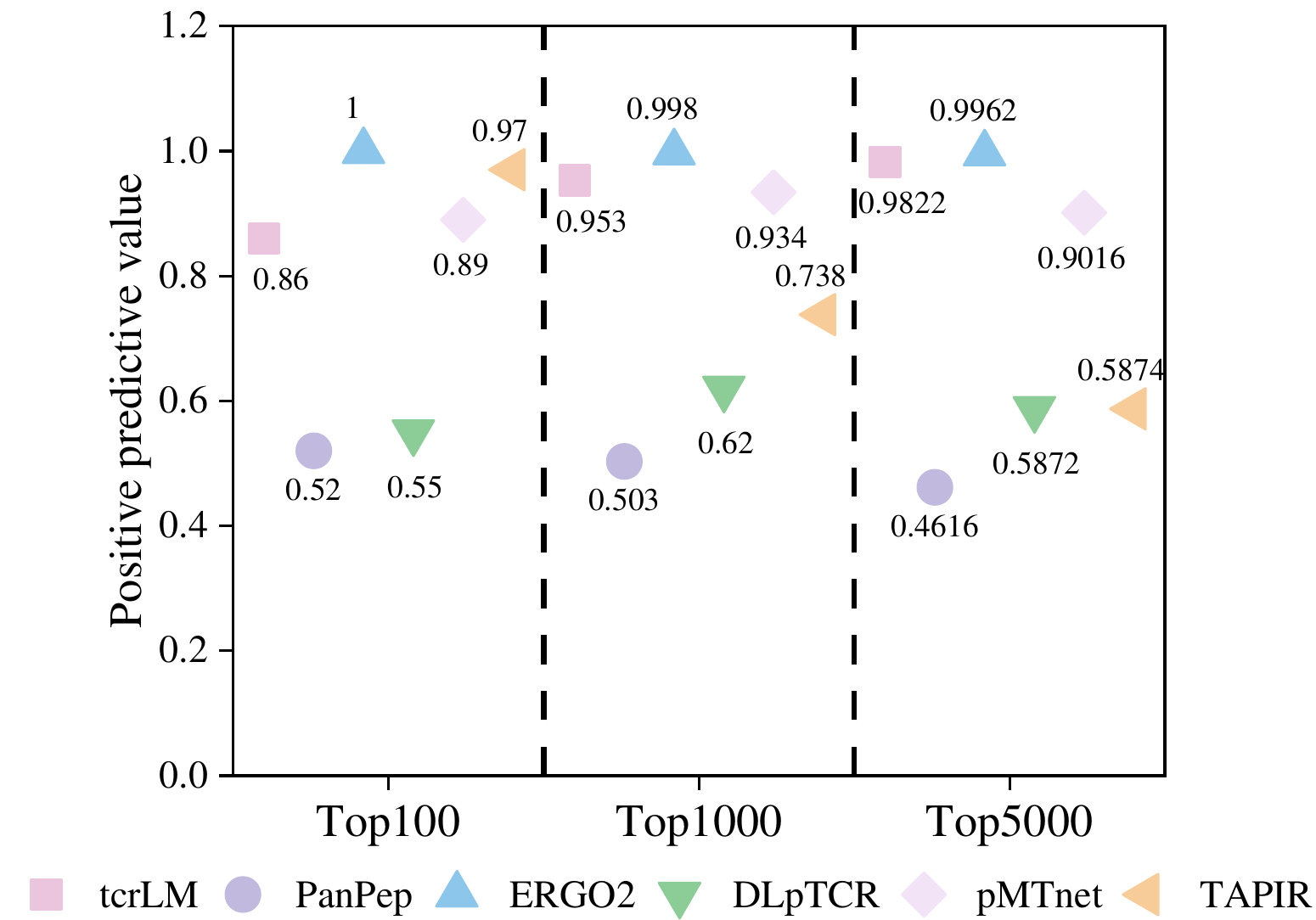}}
                  \subfloat[PPV on COVID-19 set]{\includegraphics[width=0.5\linewidth]{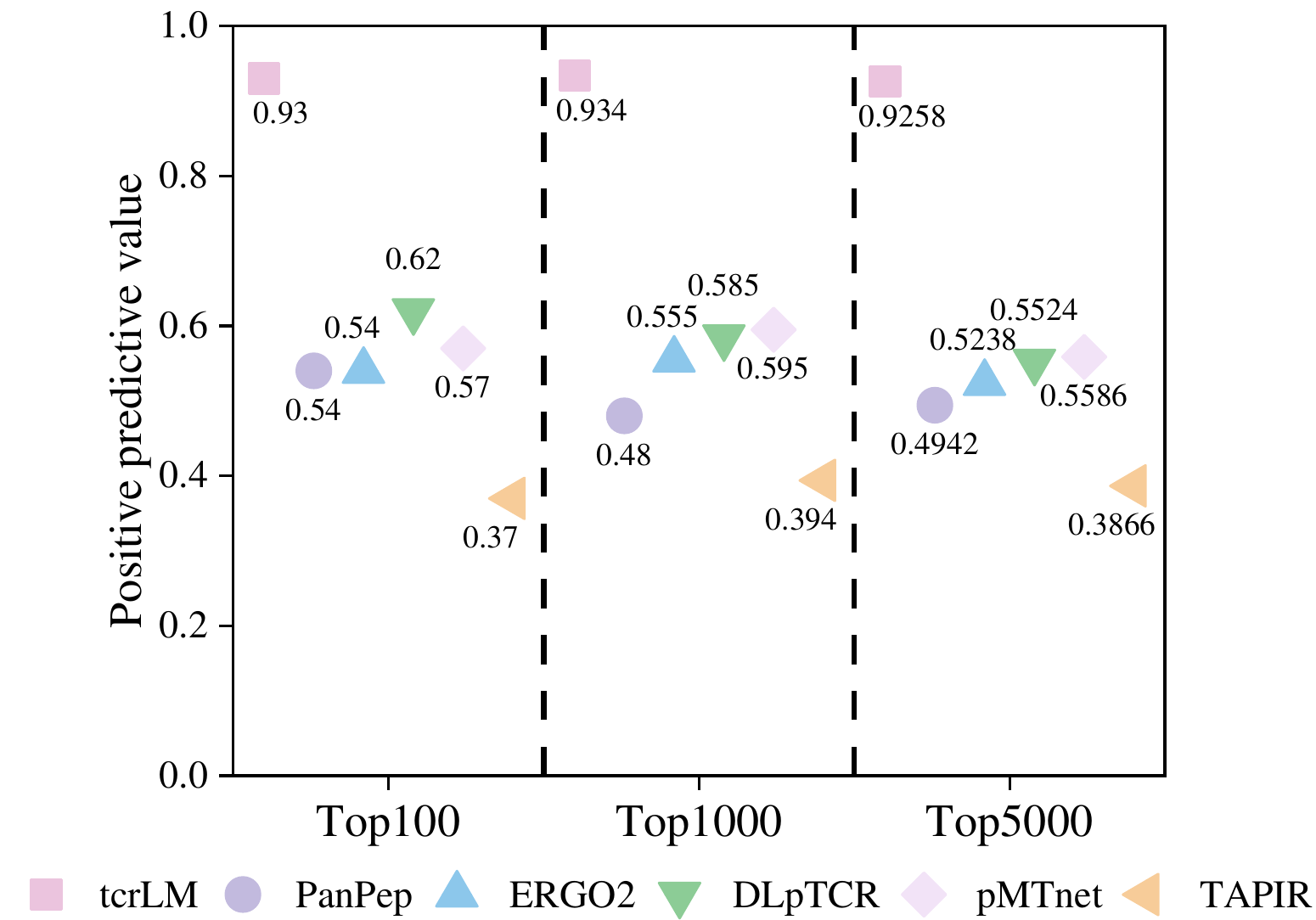}}
    \end{minipage}
        \caption{Performance evaluation of tcrLM and four comparative methods on external and COVID-19 test set. (a-b) Performance metrics of tcrLM and competing methods on external and COVID-19 test set, respectively. (c-d) Positive predictive value (PPV) for the top 100, top 1000, and top 5000 predicted samples on external and COVID-19 test sets, respectively.}
    \label{fig:external}
\end{figure*}

To further evaluate the performance of our model, we constructed an external test set collected from numbers of publications. The external set includes 63,324 pTCR binding samples across 890 unique peptides and 53,439 CDR3 sequences. It is worth noting that this external set did not contain any shared peptide with our established training set, which allows us to assess the predictive capacity of tcrLM toward real-world scenarios beyond the hold-out independent test set. To create a class-balance test set, we generated an equal number of negative samples by randomly mismatching the peptides and TCRs. For objective performance evaluation, we again compared tcrLM with the five previously published methods, including PanPep~\cite{gao2023pan}, ERGO2~\cite{springer2020prediction}, pMTnet ~\cite{lu2021deep}, DLpTCR~\cite{xu2021dlptcr} and TAPIR~\cite{fast2023tapir}. As shown in Fig.~\ref{fig:external}(a), tcrLM achieved AUROC and AUPR values exceeding 0.9, and significantly outperformed the second-best method, ERGO2, whose AUROC and AUPR values were only about 0.7. Furthermore, we calculated the positive predictive value (PPV) of the top 100, top 1000, and top 5000 predictions, and the results indicated significant differences in the PPV performance across these methods on the external set (Fig.~\ref{fig:external}c). Our method, along with ERGO2 and pMTnet, achieved high PPV values, whereas PanPen, DLpTCR and TAPIR performed poorly. We further explored the reason and found that the training datasets of ERGO2 and pMTnet overlapped with the external test set, which likely explains their relatively high performance on PPV metric. In contrast, our training set did not include any sample from the external test set, demonstrating that our method achieved superior generalizability.

We further tested tcrLM's capacity to predict bindings between virus-derived antigens and human TCRs. We collected a total of 520,000 bindings between COVID-19 virus-derived antigens and human TCRs from the ImmuneCODE database~\cite{nolan2020large}. For class balance, we generated an equal number of negative samples by randomly mismatching the peptides and TCRs. As a result, we create a million-scale COVID-19 test set, which is the largest pTCR binding test set to date. In particular, this test set did not contain any samples included in the training set. The performance evaluation results showed that tcrLM achieved AUROC and AUPR values of 0.595 and 0.602 (Fig.\ref{fig:external}b, respectively. In contrast, the competing methods obtained AUROC and AUPR values only slightly above 0.5, which is nearly close to random guessing. For the PPV metric, our model consistently achieved 93\% PPV value, significantly better than all competitors, whose PPV values remained below 60\% (Fig.\ref{fig:external}d). Overall, the significant advantage over previous methods strongly validated the robust generalizability of tcrLM and highlights its potential in facilitating effective immune-based therapies and vaccine design targeting the COVID-19 virus.

\subsection{tcrLM outperforms other protein language models}
\begin{figure*}[!htbp]
    \centering
    \begin{minipage}[t]{1\linewidth}
        \centering
         \subfloat[Performance on external set]{\includegraphics[width=1\linewidth]{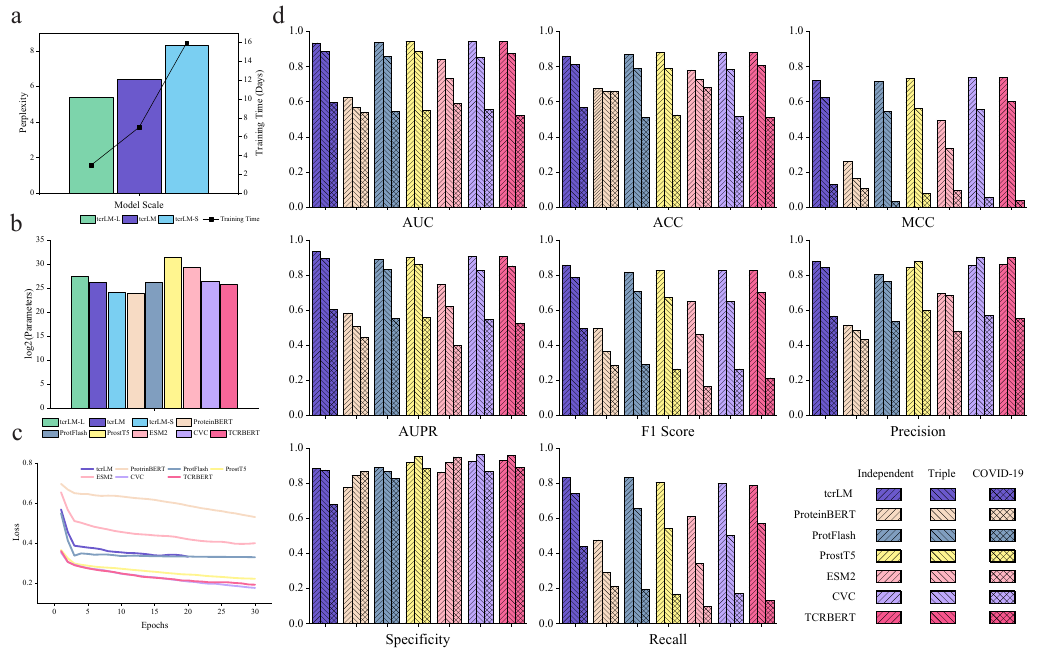}}
    \end{minipage}
        \caption{Comprehensive evaluation of tcrLM and other four protein language models. (a) Perplexity and training time of three tcrLM variants on our established large-scale dataset comprising over 100 million TCR CDR3 sequences. (b) Comparison of parameter sizes of three tcrLM variants and four other protein language models. (c) Loss curves of tcrLM-M and competing protein language models. (d) Performance comparison of tcrLM-M and competing protein language models on independent, external and COVID-19 test sets.}
    \label{fig:llm}
\end{figure*}

To explore the influence of parameter size on performance, we built three variants of the tcrLM model: tcrLM-S, tcrL-M and tcrLM-L, corresponding to small, medium, and large parameter sizes, respectively. Detailed configurations of the tcrLM variants were summarized in Table~S1. We pre-trained them on our collected 100M-scale TCR CDR3 sequences, and the three variants spent approximate 3, 7 and 16 days on our workstation, respectively. We found that tcrLM-L achieved lowest perplexity (Fig.~\ref{fig:llm}a), suggesting that a large number of parameters improved the model's ability to fit sequence data effectively. 

We further fine-tuned four other protein language models pretrained on generic protein sequences set (e.g. UniRef50 and UniRef90) to predict pTCR bindings. These language models include proteinBERT~\cite{brandes2022proteinbert},  ProtFlash~\cite{wang2023deciphering}, ProstT5~\cite{heinzinger2024bilingual}, ESM2~\cite{lin2023evolutionary}, CVC~\cite{goldner2024self} and TCR-BERT~\cite{wu2024tcr}. We compared their parameter sizes (Fig.~\ref{fig:llm}b, $\log_2$ scale), and found that ProstT5 had the largest parameter size (3 billion), while tcrLM-S and ProteinBERT had much smaller parameter sizes. We used tcrLM-M in performance evaluation experiments, because it had slightly more parameters than ProteinBERT but much fewer than the other protein language models.

For pTCR binding prediction, we fed the TCR and antigen embeddings derived from these pretrained language models into the projection head. During fine-tuning stage, the parameters of the pretrained encoders were kept frozen. To evaluate the degree of fitness to the pTCR training samples, we plotted their loss curves over training epochs (Fig.~\ref{fig:llm}c). The results demonstrated that all models converged, with CVC achieving the lowest loss value, followed by TCR-BERT and ProstT5. Next, we evaluated their performance on the independent, external and COVID-19 test sets, respectively. The results showed that tcrLM achieved superior performance and generalization capability on most metrics across all three test sets (Fig.~\ref{fig:llm}d). Although ProstT5 surpassed tcrLM on certain performance metrics, this came at the cost of a parameter size approximately 40 times larger than that of tcrLM. CVC and TCR-BERT demonstrate comparable performance to tcrLM on independent and external test sets. However, their performance on the COVID-19 test set falls short of tcrLM, indicating slightly inferior generalization capabilities. Other models exhibited inferior performance compared to tcrLM, underscoring tcrLM's ability to outperform previously published protein language models in TCR-specific tasks.

\subsection{Predicted TCR-antigen bindings indicates immunotherapy outcomes}
To further explore the predictive capacity of tcrLM, we conducted an exploratory analysis of the predicted pTCR binding scores on a cohort of patients with advanced melanoma~\cite{RIAZ2017934}. This cohort consists of 29 patients who underwent treatment by immune checkpoint inhibitors. Based on the TCR-seq and genomic sequencing data, we took each missense mutation as an anchor, and generated all possible 9-mer peptides covering this anchor. After extracting CDR3 sequences from the TCR-seq data, we created all conceivable peptide-CDR3 pairs for each patient, yielding a total of 81,851,486 pTCR candidates.

\begin{figure}[htbp]
    \centering
    \begin{minipage}[b]{0.45\textwidth}
        \centering
         \subfloat[]{\includegraphics[width=\linewidth]{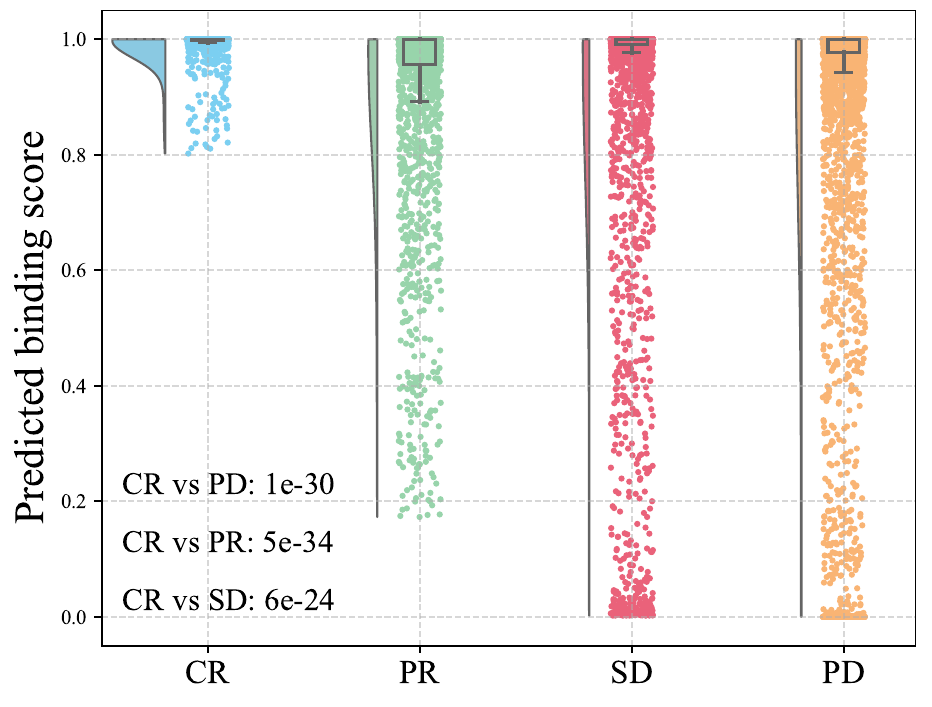}}
    \end{minipage}
    \begin{minipage}[b]{0.45\textwidth}
        \centering
         \subfloat[]{\includegraphics[width=\linewidth]{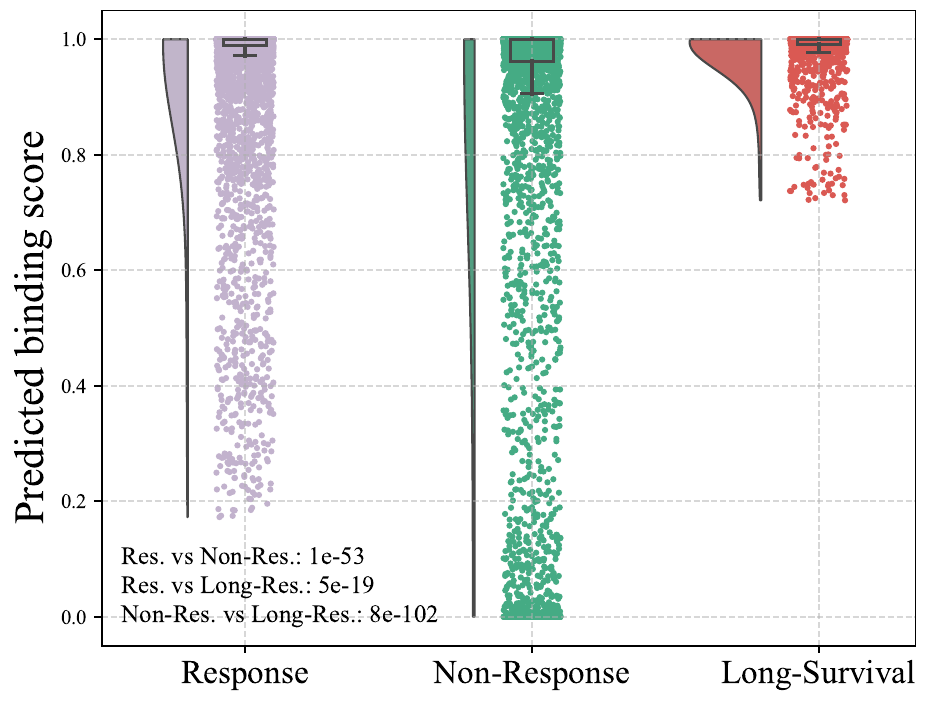}}
    \end{minipage}\\
    \begin{minipage}[b]{0.46\textwidth}
        \centering
         \subfloat[]{\includegraphics[width=\linewidth]{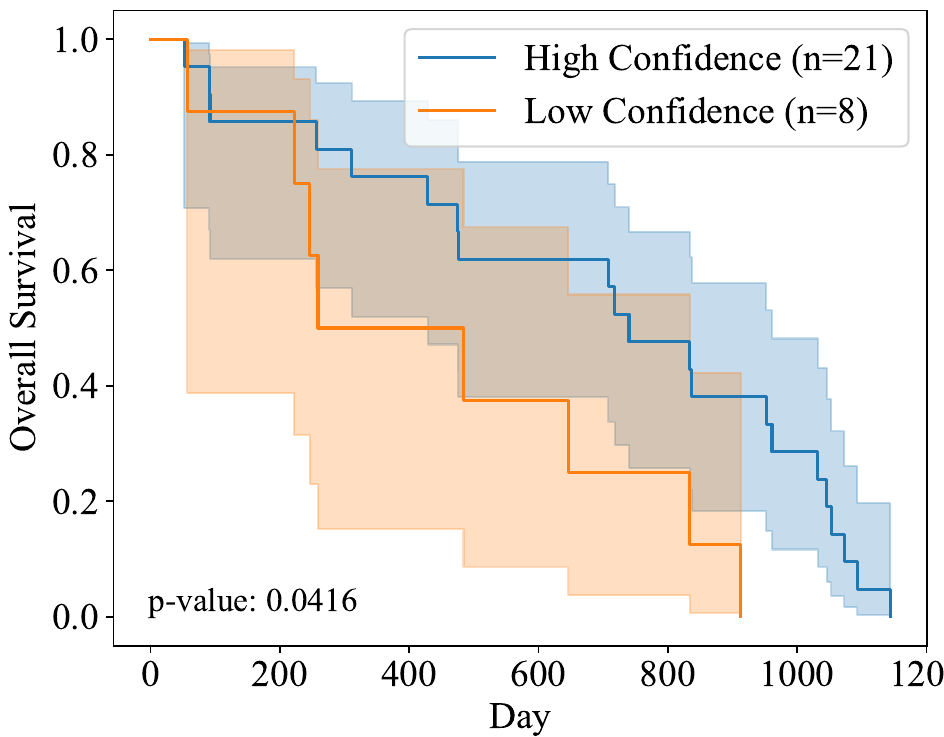}}
    \end{minipage}
    \begin{minipage}[b]{0.45\textwidth}
        \centering
         \subfloat[]{\includegraphics[width=\linewidth]{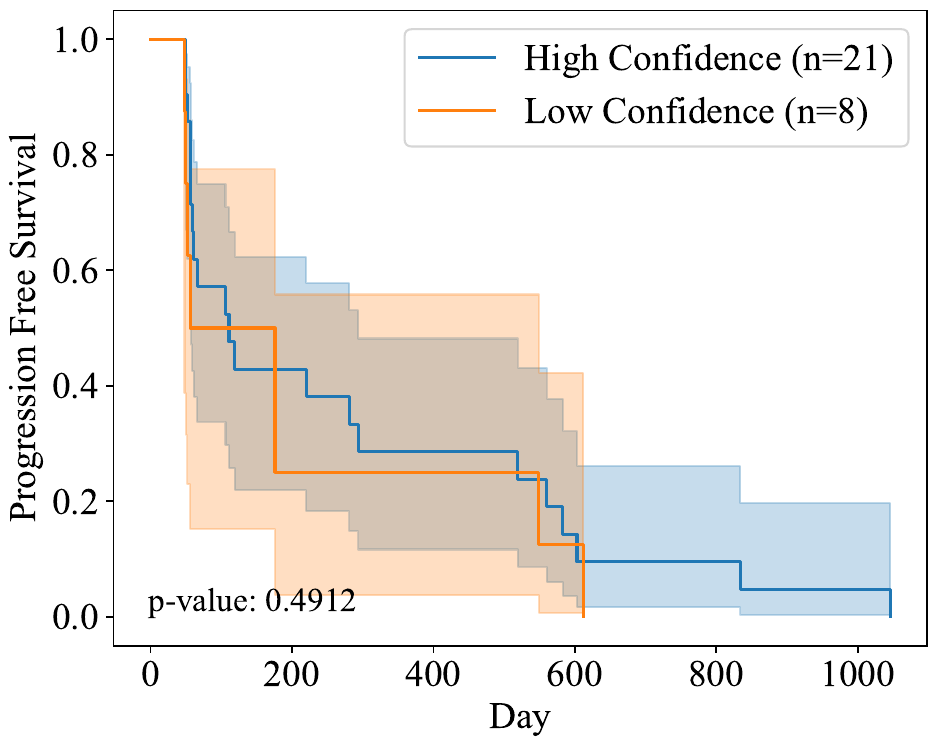}}
    \end{minipage}
    \caption{Correlation between predicted pTCR scores and immunotherapy outcomes in a melanoma cohort. (a-b) Violin and boxplots of predicted pTCR binding scores regarding the different immunotherapy response groups (p-value<0.01, $F$-tests). (c-d) Overall survival and progression free survival curves between stratified patient groups with high- and low-confidence pTCR bindings.}
    \label{fig:imunotherapy}
\end{figure}

We scored these candidate pairs using fine-tuned tcrLM, and selected the top 25,000 highest-scored pTCRs for each patient. By stratifying the patients into benefit, non-benefit, and long-term survival groups, we observed that patients in the long-term survival group had higher scored pTCRs compared to the other groups (Fig.~\ref{fig:imunotherapy}a). Subsequently, we categorized the patients into four categories according to RECIST criteria: Complete Remission (CR, $n$=2), Partial Remission (PR, $n$=5), Stable Disease (SD, $n$=9), and Progressive Disease (PD, $n$=12) groups, and found the predicted bindings scores of four groups differed significantly from each other (Fig.~\ref{fig:imunotherapy}b). The analysis of variance (ANOVA) verified the statistical differences among these groups ($F$-test, p-value<0.01). This findings validated the correlation between higher pTCR predicted scores and better immunotherapy outcomes, offering a new perspective and a possible clinical indicator for tumor immunotherapy. 

Finally, to confirm the correlation between highly scored pTCR pairs by tcrLM and improved clinical outcomes, we conducted survival analysis on this cohorts. We considered the top 2\% highly-scored pTCR pairs as high-confidence bindings, and stratified the patients with such pairs into the high-confidence group, while the remaining patients were placed into the low-confidence group. The survival analysis showed that the high-confidence group exhibited significantly higher overall survival (OS) and progression free survival (PFS) compared to the low-confidence group (Fig.~\ref{fig:imunotherapy}c-d). Although the difference in PFS did not exhibited statistical significance, it may be attributed to the limited sample size of the cohort. These findings suggest that the patients with highly scored pTCR bindings benefited more from immunotherapy and yielded favourable clinical outcomes.

\subsection{tcrLM captures residual biochemical properties and positional preference }
\begin{figure*}
    \centering
    \begin{minipage}{0.46\textwidth}
        \centering
        \subfloat[]{\includegraphics[width=\linewidth]{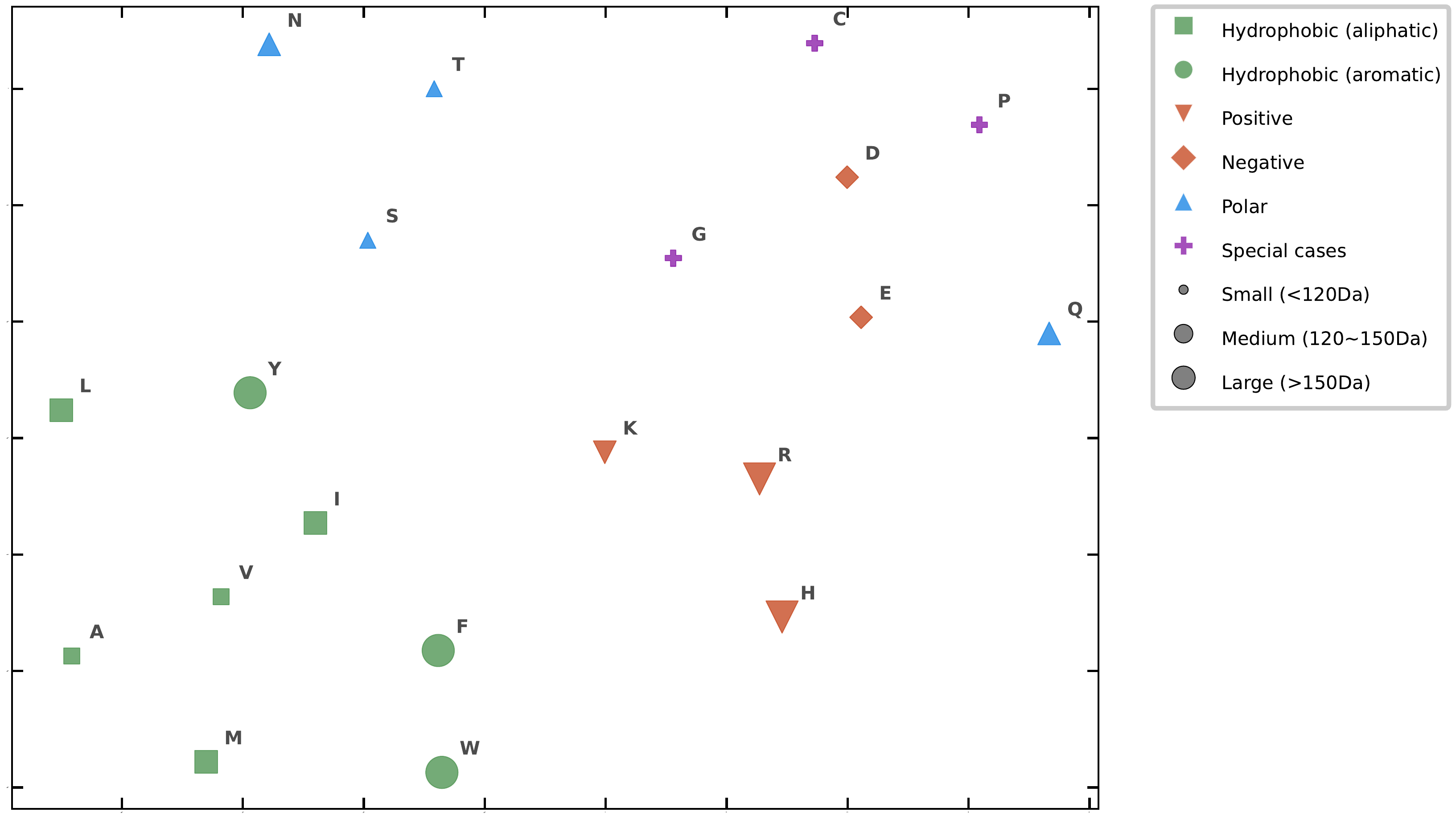}}
    \end{minipage}  
    \begin{minipage}[]{0.53\textwidth}
        \centering
        \subfloat[]{\includegraphics[width=\linewidth]{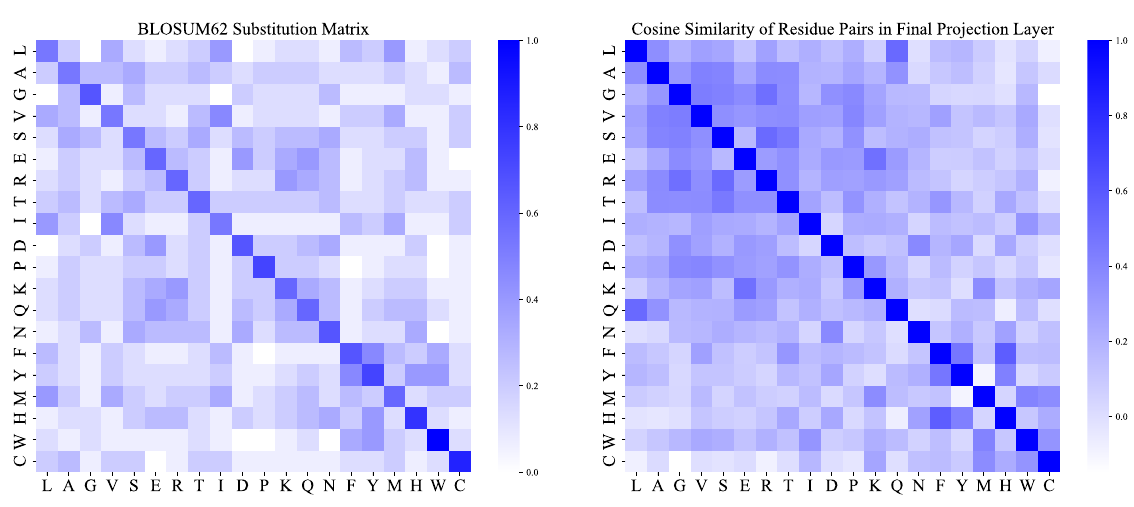}}
    \end{minipage} \\
    \vspace{0.3cm}
    \begin{minipage}[]{0.32\textwidth}
        \centering
        \subfloat[]{\includegraphics[width=\linewidth]{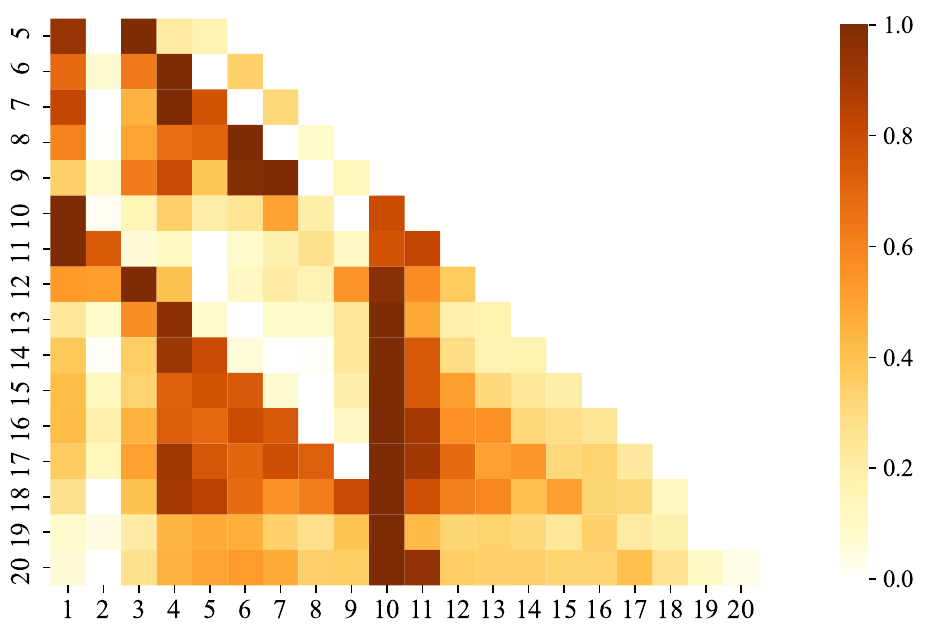}}
    \end{minipage} 
     \begin{minipage}[]{0.65\textwidth}
        \centering
        \subfloat[]{\includegraphics[width=\linewidth]{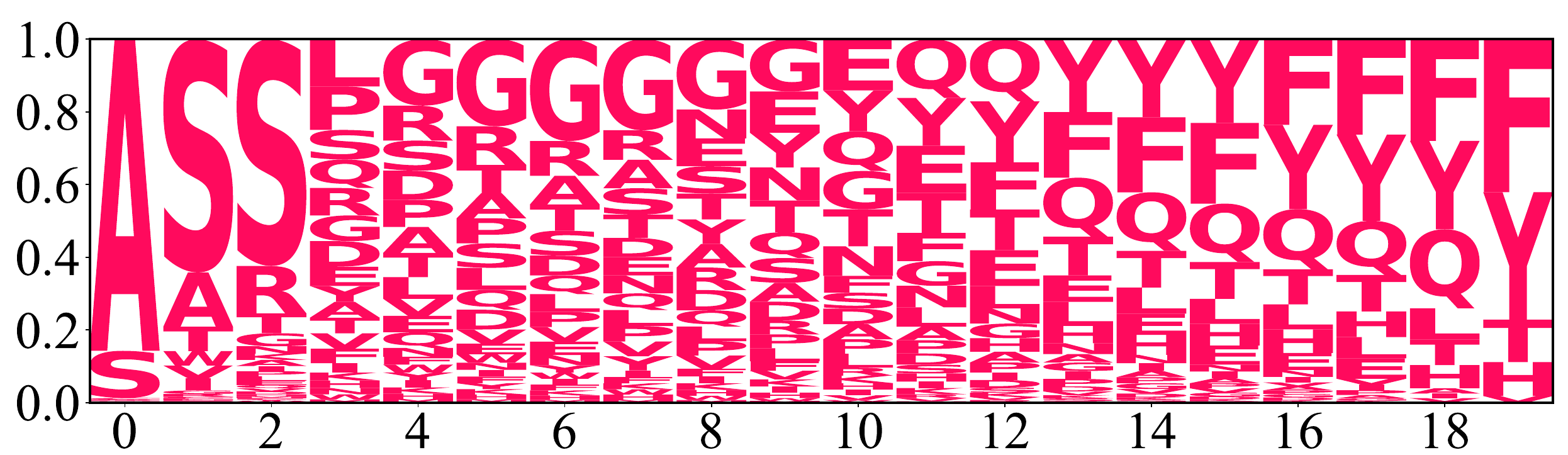}}
    \end{minipage} 
    \caption{tcrLM captured the biochemical properties and positional preference of amino acids within TCR sequences. (a) tSNE plot of the high-dimensional embeddings of 20 amino acids obtained from the stacked encoder. (b) Heatmaps of the computed cosine similarities using the BLOSUM62 substitution matrix and learned representations of of 20 amino acids by tcrLM. (c) Heatmap generated by cumulative attention weights regarding each position for 5-21mer TCR sequences. (d) Sequence logo of the CDR3 sequences where the relative importance of amino acids at each position is quantified by the cumulative attention weights.}
        \label{fig:t-SNE}
\end{figure*}

To investigate whether tcrLM has learned the physicochemical properties of amino acids, we applied t-distributed stochastic neighbor embedding (t-SNE)~\cite{2008Visualizing} to project the high-dimensional embeddings of 20 amino acids obtained from the attention weight matrices into a two-dimensional space. Similar to the observations in other protein language models such as ESM-1b~\cite{rives2021biological} and ProtFlash~\cite{wang2023deciphering}, the t-SNE plots reveal distinct clustering patterns of amino acids within the embedding space (Figure~\ref{fig:t-SNE}a). Specifically, amino acids with similar physicochemical properties formed well-defined clusters: hydrophobic and polar residues were distinctly grouped, and aromatic residues are concentrated in another cluster. Furthermore, amino acids are also grouped according to molecular weight and charge. These results highlights the tcrLM’s capability to learn and encode fundamental amino acid properties in the latent space. To further evaluate the model's ability to capture relationships between different amino acids, we calculated the cosine similarity for each residue pair based on their embeddings from the final projection layer, and displayed the results in a heatmap (Figure~\ref{fig:t-SNE}b). For comparison, we presents the heatmap of the BLOSUM62 substitution matrix, a well-established benchmark derived from expert-curated sequence alignments. Notably, a strong concordance was observed between the cosine similarity scores and the BLOSUM62 substitution scores. The Pearson correlation coefficient and mean squared error (MSE) between the two heatmaps were 0.692 and 0.0027, respectively. These findings indicate that tcrLM effectively captures the likelihood of similar amino acids substituting for one another, aligning with biologically ``tolerant" substitutions that minimally affect protein structure and function. 

To elucidate the importance and positional preferences of amino acids within TCR CDR3 sequences, we examined the cumulative attention weights at each position across sequences of varying lengths (5–20 residues), and visualized the results using a heatmap (Figure~\ref{fig:t-SNE}c). For shorter CDR3 sequences (\(\text{len} < 9\)), amino acids located at the first position and the third-to-last position were found to be critical for pTCR binding. In contrast, for relatively long CDR3 sequences (\(\text{len} > 10\)), the 10th position exhibited a markedly higher contribution, underscoring its significance in binding to antigens. We further examined the  importance of amino acids at each position using cumulative attention weights (Figure~\ref{fig:t-SNE}d), and revealed that the amino acid \( \text{A} \) at the first position, and \( \text{S} \) at the second and third are highly significant. Moreover, \( \text{F} \) and \( \text{Y} \) showed predominantly importance in the tail regions of the CDR3 sequence, highlighting their critical role in antigen binding.

\section{Conclusion}

This study proposed a BERT-based large language model, named tcrLM, to enhance the accuracy of predicting pTCR binding specificity by utilizing the strong capability in feature extraction of large language models. We have pre-trained the model on more than 100 million TCR sequence, and verified its performance on downstream tasks. Our experimental results demonstrated that tcrLM exhibited superior performance compared to the current four state-of-the-art methods on independent, external and COVID-19 test sets. Moreover,   tcrLM’s predicted TCR-neoantigen binding scores is indicative of the immunotherapy responses and clinical outcomes in a melanoma cohort. These findings highlight tcrLM's potential applicability in vaccine design and immunotherapy treatment.

While several studies have proposed TCR-specific language models for predicting pTCR binding, these models exhibited limited performance on our curated test set. This limitation likely arises from their pretraining on relatively small datasets of TCR CDR3 sequences, which restricts their ability to learn the meaningful representations. In contrast, we established a large-scale dataset comprising over 100 million CDR3 sequences, which enables tcrLM to capture richer and more nuanced representations of TCR CDR3 sequences, thereby achieving enhanced performance in downstream prediction tasks. Additionally, we fine-tuned other protein language models with large parameter sizes  for pTCR binding prediction, and compared their performance to our lightweight model. Remarkably, despite its smaller parameter size, our model achieved comparable or even superior performance to these larger language models. These findings demonstrate that a lightweight model pretrained merely on TCR CDR3 sequences can achieve competitive predictive capabilities in pTCR binding specificity.

When utilizing the predicted pTCR binding scores to study immunotherapy responses, the absence of patient-specific HLA alleles make us to consider all possible neoantigens and match them with TCRs to generate candidate pTCR pairs. In fact, most neoantigens cannot specifically bind to MHC class I molecules for presentation on the cell surface and, consequently, cannot be recognized by TCRs. Despite this discrepancy from actual biological processes, the predicted pTCR binding scores exhibited a significant correlation with immunotherapy responses and patients’ overall survival. These findings indicate that our model effectively captures the global landscape of pTCR interactions and demonstrates robust predictive power for immunotherapy outcomes.

\section{Materials and Methods}

\subsection*{Data source and preprocessing}
For language model pre-training, we set about to establish a largest set of TCR CDR3 sequences. For this purpose, we collected 113,888,692 distinct TCR CDR3 sequences from more than ten databases and publications (Table S2). With over 100 million sequences, our model is able to capture underlying feature from TCR sequences. The sequence frequency distribution with respect to lengths is illustrated in Fig.S1(a), which reveals that most CDR3 sequences range between 10 to 20 amino acids in length. The frequency distribution of associated antigen sequences is shown in Fig.S1(b), indicating that most antigens 9mer in length.

For fine-tuning, we build a benchmark dataset of pTCR bindings by collecting a substantial amount of pTCR binding data from various databases. We consider both the $\alpha$ and $\beta$ chains of TCR and treat them as independent CDR3 sequences, as previous studies have demonstrated both of them play critical role in antigen recognition. The pTCR dataset consists of 109,554 bindings, covering 1,377 unique antigens and 104,623 unique TCR CDR3 sequences. To our best knowledge, this the largest pTCR binding dataset to date.

For further performance validation, we establish an external test set comprising  63,324 pTCR bindings across 890 unique peptides and 53,439 CDR3 sequences collected from numbers of publications. In addition, a large-scale test set are sourced from ImmuneCODE database~\cite{nolan2020large}, consisting of 520,000 pTCR bindings between COVID-19 virus-derived antigens and human TCRs. 

The negative pTCR samples were created by randomly mismatched TCR and peptide sequences, till the same number of negative samples were generated. During the fine-tuning stage, we randomly select 10\% of the pTCR binding dataset as an independent test set, while the remaining 90\% is used as the training set. The five-fold cross-validation is used to optimize the model hyperparameters

\subsection*{TcrLM architecture}
The language model is trained to recover the masked segment using the surrounding sequences on the collected more than 100 million TCR sequences. As illustrated in Fig.~\ref{fig:model}(a), each TCR CDR3 sequence is padded to a fixed length of 20, since all CDR3 sequences do not exceed 20. We randomly generate a mask of length between 3 to 5 to cover a continuous segment in the TCR sequence. The masked sequence passes through an embedding layer that converts each token into a high-dimensional embedding. The embeddings are then taken as input to the encoder and ultimately mapped into a 20x512 matrix, which is transformed to 20*21 matrix via a linear layer. We introduced the rotary positional embedding (RoPE)\cite{su2024roformer} into tcrLM. This method encodes absolute positions using rotation matrices and integrates explicit relative positional dependencies into self-attention calculations. RoPE offers the flexibility in sequence length, decays the dependencies between tokens as their relative distance increases, and provides the capability to enhance linear self-attention with relative positional encoding. Furthermore, we incorporated the mixed chunk attention (MCA) \cite{hua2022transformer} to combine the advantages of both local attention and linear attention mechanisms.

 In our practice, we applied different learning rates to distinct components of the model. For the encoder, the learning rate was set to 1e-6, while for the subsequent fully connected layers, it was adjusted to 1e-4. The optimizer was set to Adam. The pre-training was conducted on a CentOS Linux 8.2.2004 (Core) system, equipped with an Intel(R) Xeon(R) Silver 4210R CPU running at 2.40GHz, along with a GeForce RTX 4090 GPU and 128GB of RAM. The model was implemented using the PyTorch 2.2.1 framework. 

\subsection*{Predicting pTCR binding}
The pre-trained encoder is subsequently used for the prediction task of pTCR binding. As shown in Fig.~\ref{fig:model}(b), the antigen sequences and TCR sequences are padded to the maximum length of 15 and 20, respectively. After passing through an embedding layer, these two sequences are taken as input of the pre-trained encoder, and transformed into matrices of dimensions 15x512 and 20x512. Subsequently, these two matrices are concatenated along the feature dimension to form a 35x512 matrix, which is then flattened into a 17,920-dimensional vector. This vector passes through a fully-connected layer and is projected to a 2-dimensional vector, followed by softmax layer to obtain the final predicted probabilities. During the fine-tuning process, the pre-trained encoder is frozen. For pTCR binding prediction task, we use the cross-entropy as the loss function:
\begin{equation}
\mathcal{L}_{cls} = -\frac{1}{N} \sum_{i=1}^N \left[ y_i \cdot \log(p_i) + (1 - y_i) \cdot \log(1 - p_i) \right]
\end{equation}
in which $y_i$ and $p_i$ represent the actual and predicted pTCR binding, and $N$ is the total number of training samples.

\subsection*{Virtual adversarial training}
Given the vast diversity of the TCR repertoire, the training data currently available is still limited and even biased. This poses a tough challenge on the development of robust prediction models for pTCR binding. To address this challenge, we employs virtual adversarial training to enhance the model's generalizability. Since protein sequence are discrete, we introduce perturbations in the embedding layer of the sequences~\cite{miyato2016adversarial}. For this purpose, we create adversarial examples designed to maximize the loss function by applying adversarial perturbations to the sequence embeddings, namely, the adversaries are generated in the direction of gradient ascend and constrained by L2 norm. Formally, we define the adversarial loss as below:
\begin{equation}
    \begin{aligned}
    L_{\mathrm{vadv}}(x,\theta)&=D\left[p(y|x,\hat{\theta}),p(y|x+r_{\mathrm{vadv}},\theta)\right]),
\\\text{ where }
r_{\mathrm{vadv}}&=\arg\max_{r;\|r\|\leq\epsilon}D\left[p(y|x_*,\hat{\theta}),p(y|x+r)\right], \end{aligned}
\end{equation}
$D$ represents the function that measures the divergence between two distributions, $p(y|x)$ denotes the probability of the model predicting label $y$ given input $x$, $r_{\mathrm{vadv}}$ is a virtual adversarial perturbation regarding the input sample $x$. This perturbation strives to maximize the divergence between $p(y|x_*,\hat{\theta})$ and $p(y|x+r)$ by choosing the direction of gradient ascent.

Virtual adversarial training requires the model to minimize not only the risk on actual observed data, but also the risk arising from adversarial loss. This approach helps reduce the model's sensitivity to slight input variations, thereby enhancing its generalizability. By incorporating virtual adversarial training, we aim to build a more robust and reliable prediction model, capable of handling the vast diversity of the TCR repertoire even with limited and potentially biased training data.





\bibliography{sn-bibliography}

\end{document}